\documentclass{article}
\usepackage{times}
\usepackage{amsfonts}
\usepackage{graphicx}
\usepackage[pdfmark]{hyperref}
\begin{document}
\noindent
{\Large  A NOTE ON THE QUANTUM OF TIME}
\vskip1cm
\noindent
{\bf Jos\'e M. Isidro}\footnote{\tt joissan@mat.upv.es}, {\bf J.L. Gonz\'alez--Santander}\footnote{\tt jlgonzalez@mat.upv.es} and {\bf P. Fern\'andez de C\'ordoba}\footnote{\tt pfernandez@mat.upv.es}\\
Grupo de Modelizaci\'on Interdisciplinar, Instituto de Matem\'atica Pura y Aplicada,\\ Universidad Polit\'ecnica de Valencia, Valencia 46022, Spain
\vskip1cm
\noindent
{\bf Abstract}   Quantum mechanics rests on the assumption that time is a classical variable. As such, classical time is assumed to be measurable with infinite accuracy. However,  all real clocks are subject to quantum fluctuations, which leads to the existence of a nonzero uncertainty in the time variable.  The existence of a quantum of time modifies the Heisenberg evolution equation for observables. In this letter we propose and analyse a generalisation of Heisenberg's equation for observables evolving in real time (the time variable measured by real clocks), that takes the existence of a quantum of time into account. This generalisation of Heisenberg's equation turns out to be a delay--differential equation.

\section{Introduction}

In its usual formulation, quantum mechanics relies on the idealisation that all measuring devices are perfectly classical apparatuses, not subject to quantum fluctuations. This is however not true, as everything within the Universe,  measuring devices included,  is subject to some level of quantum fluctuations. When measuring spacetime, this statement implies that neither clocks nor rulers can be perfectly classical. Rather, they are subject to limitations on their accuracy; one cannot measure space and time beyond a minimum level of uncertainty.  Rulers as measuring devices and the corresponding uncertainties in the determination of space variables have been analysed in \cite{SPACE}. In this letter we will concentrate on clocks as measuring devices that are themselves also subject to the laws of quantum mechanics. The authors of \cite{EGUSQUIZA, GPUNO} call these apparatuses {\it real}\/ clocks, as opposed to {\it classical}\/ clocks. Classical time, the variable measured by a classical clock and denoted $t$, is not subject to any uncertainty. Real time, denoted $T$, is the physical variable measured by a real clock; it is to $T$ that the quantum of time applies.

One important implication of the existence of a quantum of time is that the evolution equations of quantum mechanics, when written in terms of real time $T$, pick up additional terms with respect to the corresponding equations when written in terms of the classical time variable $t$.  These additional terms spoil unitarity and lead to decoherence effects \cite{HAUER}. 

This letter is devoted to analysing the consequences of such effects on the usual Heisenberg evolution equation, under a Hamiltonian $H$, for observables $O$ that do not to depend explicitly on time, 
\begin{equation}
\frac{\partial O}{\partial t}={\rm i}\left[H,O\right].
\label{quatt}
\end{equation}
We first set the stage in section \ref{deco} with a brief summary of  \cite{EGUSQUIZA, GPUNO}. Our generalisation of eqn. (\ref{quatt}) is presented in section  \ref{heys}. We round up in section \ref{kommentar} with a discussion on the possible uses of delay--differential equations in quantum gravity.

\section{Decoherence in the course of time}\label{deco}

Let a certain Hamiltonian $H$ be given to generate translations along $t$, and let $U(t)$ be the corresponding unitary evolution operator. Call $P_t(T)$ the probability that the resulting measurement of the clock variable $T$ correspond to the value $t$. Given the density matrix $\rho$ for a system under consideration in Schroedinger's picture, in Heisenberg's picture we have a density matrix $\rho(T)$ 
\begin{equation}
\rho(T)=\int_{-\infty}^{\infty}{\rm d}t\,U(t)\rho U(t)^{\dagger}P_t(T).
\label{uno}
\end{equation}
Unitarity is lost because $\rho(T)$ is a superposition of density matrices associated with different $t$'s, each one of which evolves unitarily. Further assume that the real clock is semiclassical, so  $P_t(T)$ can be set equal to $f(T-T_{\rm max}(t))$, with $f$ a function decaying very rapidly for values of $t$ away from the maximum $T_{\rm max}$ of the probability distribution function. Then to leading order ({\it i.e.}, in a semiclassical analysis) one finds \cite{EGUSQUIZA, GPUNO}
\begin{equation}
\frac{\partial\rho(T)}{\partial T}={\rm i}[\rho(T), H]+\sigma(T)[H,[H,\rho(T)]],
\label{dos}
\end{equation}
where $\sigma(T)$ is the rate of change of the width of the distribution $f(T-T_{\rm max}(t))$. An estimate for the function $\sigma(T)$ is 
\begin{equation}
\sigma(T)=\left(\frac{T_{\rm Planck}}{T_{\rm max}-T}\right)^{1/3}T_{\rm Planck},
\label{plann}
\end{equation}
where $T_{\rm Planck}=10^{-44}$ seconds is Planck's time. Integrating (\ref{dos}) one finds the evolution of the density matrix in the energy eigenbasis:
\begin{equation}
\rho(T)_{nm}=\rho(0)_{nm}\exp\left(-{\rm i}\omega_{nm}T\right)\exp\left(-\omega^2_{nm}T_{\rm Planck}^{4/3}T^{2/3}\right).
\label{tres}
\end{equation}
A pure state will inevitably become a mixed state due to the real exponential on the right--hand side; ultimately this stems from the impossibility of having a perfectly classical clock.

\section{A time--delayed Heisenberg equation}\label{heys}

A computation shows that one can recast the evolution equation (\ref{dos}) as follows:
\begin{equation}
\frac{\partial\rho}{\partial T}={\rm i}\left[\tilde\rho, H\right], \qquad \tilde\rho:=\rho(T-\sigma)=\exp\left(-\sigma\frac{\partial}{\partial T}\right)\rho(T).
\label{dumbo}
\end{equation}
We have made use of the fact that ${\rm i}[\rho,H]=\partial_t\rho$, which we can further approximate as $\partial_T\rho$ in the semiclassical regime.
We remark that the tilde on the right--hand side is absent from the left--hand side. The nonzero quantum of time causes a backward time shift within $\rho$. This shift  vanishes as $\sigma\to 0$, {\it i.e.}, as our clock becomes classical. Additional dependence on space variables (not considered here) will produce further decoherence effects  \cite{SPACE}.

Inspired by the previous reasoning we propose that an arbitrary observable $O$ must evolve semiclassically under real time $T$ as governed by the equation
\begin{equation}
\frac{\partial O}{\partial T}={\rm i}\left[H,\tilde O\right], \qquad \tilde O:=O(T-\sigma)=\exp\left(-\sigma\frac{\partial}{\partial T}\right)O(T).
\label{cinq}
\end{equation}
Eqn. (\ref{cinq}) is best understood as a delay--differential equation \cite{HALE}: the right--hand side is delayed by $\sigma$ with respect to the left--hand side. The existence of a quantum of time causes the classical--time Heisenberg eqn. (\ref{quatt}) to become its real--time counterpart (\ref{cinq}), at least semiclassically. We further observe that, although $\exp\left(-\sigma\frac{\partial}{\partial T}\right)$ is a unitary operator (best seen by rewriting it as $\exp\left({\rm i}\sigma{\rm i}\frac{\partial}{\partial T}\right)$), the transformation from $O$ to $\tilde O$ is {\it not}\/ the unitary transformation law for operators, $O\rightarrow \exp\left(-\sigma\frac{\partial}{\partial T}\right)O\exp\left(\sigma\frac{\partial}{\partial T}\right)$ .

\section{Discussion}\label{kommentar}

Setting $O=T$ in (\ref{cinq}) we find ${\rm i}[H, \tilde T]={\bf 1}$ as usual---not quite, really, because of the tilde on top of $T$. In the presence of a quantum of time, commutators are no longer computed with their entries evaluated at equal times, as is the case in canonical quantisation. Rather, the two entries within a commutator are {\it delayed}\/ with respect to each other by $\sigma$. This is not totally unexpected. That classical gravitational fields slow down classical clocks has been known for long. Real clocks, those for which the quantum of time cannot be neglected, are also slowed down, which leads to a generalisation of the Heisenberg equation (\ref{quatt}) under the form of the delayed--differential equation (\ref{cinq}). Thus the slowdown effect on clocks becomes more apparent in quantum gravity.

One finds \cite{GPUNO} that the best accuracy $\delta T$ one can get in a measurement of the time interval $T$ is given by $\delta T=T_{\rm Planck}^{2/3}T^{1/3}$. Now the uncertainty principle for operators $A,B$ whose commutator is proportional to the identity reads $\Delta A_\psi\Delta B_\psi\geq \vert \langle\psi[A,B]\psi\rangle\vert/2$.  At least in semiclassical quantum gravity, where spacetime may still be said to have an entity of its own (albeit under the form of operator--valued coordinate functions) one is tempted to interpret the nonzero value of  $\delta T$ in terms of a nonzero commutator $[T(t_1),T(t_2)]$. We do not know what this commutator is, but we can make an educated guess by setting it to be proportional to the identity operator, antisymmetric under the exchange of $t_1$ and $t_2$, and carrying the dimensions of time squared. So our Ansatz reads 
\begin{equation}
[T(t_1), T(t_2)]=f(t_2-t_1){\bf 1}.
\label{edd}
\end{equation}
Moreover, the unknown c--number function $f$ must satisfy $f(t_1-t_2)=-f(t_2-t_1)$ and have the dimensions of time squared. Then a measurement of a real--time interval $T$ that is numerically equal to $t_2-t_1$ will saturate the inequality in the uncertainty principle for a choice of $f$ such as
\begin{equation}
f(t_2-t_1)=2\,{\rm sign}(t_2-t_1)\,T_{\rm Planck}^{4/3}(t_2-t_1)^{2/3},
\label{anss}
\end{equation}
where ${\rm sign}(t_2-t_1):=\theta(t_2-t_1)-\theta(t_1-t_2)$ is the sign function and $\theta(t)$ the Heaviside step function. Of course there are more functions $f$ and more commutators than (\ref{edd}) satisfying the necessary requirements; our (\ref{edd}) and (\ref{anss}) are the simplest choices. However our choice is particularly natural because it automatically leads to the argument of the decaying exponential in (\ref{tres}).

The reader may ask, what role does quantum gravity play here? After all, one can derive eqn. (\ref{dos}) by simply placing a clock within a thermal bath and computing fluctuations in time measurements (due to imperfections of the clock itself) with the help of Boltzmann's distribution \cite{EGUSQUIZA}. To motivate our answer we observe that one major point addressed in refs. \cite{SPACE, GPUNO} is the following question: {\it How do quantum notions, as applied to spacetime, alter our views of quantum mechanics?}\/ This point of view is complementary (in Bohr's sense of the word) to the widespread opinion that, given classical general relativity on the one hand, and quantum mechanics on the other, what remains to be done is to {\it quantise}\/ gravity---an enterprise (the quantisation of gravity) that has kept theoretical physicists busy for the last 75 years \cite{QG}. In loose terms, answering the question raised above could be seen as a step towards {\it relativising}\/ the quantum, a point of view that is dual (in Bohr's sense of the word) to {\it quantising}\/ gravity.

After completion of this work we became aware of ref. \cite{ELZE}, where issues closely related to ours are dealt with from an interesting alternative perspective.

\noindent
{\bf Acknowledgements} J.M.I. thanks Max-Planck-Institut f\"ur Gravitationsphysik, Albert-Einstein-Institut (Potsdam, Germany) for hospitality extendend over a long period of (real!) time.

\end{document}